\documentstyle[epsf,preprint,aps]{revtex}

 \def\(({\left(}
 \def\)){\right)}

\def \bea{\begin{eqnarray}}
\def \eea{\end{eqnarray}}
\def \be{\begin{equation}}
\def \ee{\end{equation}}

\def \ab2{\alpha\beta^2}

\newcommand \s {\sigma}

\begin{document}
\draft
\title{Evidence of a Critical time in Constrained Kinetic Ising models}

\author{Eduardo Follana(*) and Felix Ritort(**)}
\address{(*)  Departamento de Fisica Teorica\\
     Facultad de Ciencias,\\
    Universidad de Zaragoza,\\
     50009 Zaragoza (Spain)\\
e-mail: {\it follana@posta.unizar.es}\\ 
(**) Departamento de Matematicas,\\
Universidad Carlos III, Butarque 15\\
Legan\'es 28911, Madrid (Spain)\\
e-mail: {\it ritort@dulcinea.uc3m.es}\\}

\date{\today}
\maketitle



\begin{abstract}
We study the relaxational dynamics of the one-spin facilitated Ising
model introduced by Fredrickson and Andersen. We show the existence of a
critical time which separates an initial regime in which the relaxation
is exponentially fast and aging is absent from a regime in which
relaxation becomes slow and aging effects are present. The presence of
this fast exponential process and its associated critical time is in
agreement with some recent experimental results on fragile glasses.
\end{abstract} 

\vfill

\narrowtext
\section{Introduction}

The subject of glassy dynamics has received a lot of attention in
the last years \cite{glasses}.  During a fast enough cooling process real
glasses do reach a metastable glassy phase of free energy higher than
that of the crystal phase. Apparently the glass transition behaves as a
purely kinetic phenomenon and the glass does not equilibrate when
probed in a time scale smaller than the relaxation time. 

Laboratory experiments can measure one time extensive quantitites like
enthalpy and its associated specific heat and also the two-times
correlation function by measuring the scattering processes. These
spectra give direct information about the relaxational processes which
take place in glasses. One of the most studied relaxational processes in
glasses is the so called structural or $\alpha$ relaxation which yields
the structural relaxation time. While the $\alpha$-relaxation is a slow
process there are other faster processes which have been observed
experimentally. Close to the glass transition two fast processes have
been observed: 1) the $\beta$-relaxation process predicted by the Mode
Coupling Theory (MCT)\cite{MCT} and observed in dielectric response
measurements and 2) a faster process of order of picoseconds observed in
neutron scattering experiments \cite{neutron}. In this last case,
evidence has been reported on the existence of a critical time from the
croossover from Debye (exponential in time and diffusive in space) to
non-Debye relaxation \cite{Col}. This critical time follows a
temperature dependent Arrhenius behavior. The purpose of this work is
to show that the existence of this critical time is an essential
ingredient of some kinetic models with short range constrained dynamics.

Several types of models have been proposed to understand the dynamical
behavior of real glasses. All of them have in common the presence of a
certain type of frustration. These models can be classified in two
large classes. In the first class of models, there is frustration in the
energy function. During its dynamical evolution these systems move in
phase space avoiding configurations of higher free energy. The dynamics
can be very slow due to the existence of energy barriers (strong
metastability \cite{FrPa}) or due to entropy barriers (strong
marginality \cite{BG}). Spin glasses \cite{Books} belong to these large
class of systems where, in the most general case, disorder is not
essential and can be self-generated by the dynamics\cite{MaPaRi}.

While the first class of models (at least, in the mean-field
approximation) seem to capture the experimentally observed features
related to the (slow) $\alpha$-relaxation process and the (fast)
$\beta$-relaxation process \cite{MFSG} it is still unclear how much they
can account for this observed new type of fast Debye relaxational
process \cite{Col}. 

In the second type of models the frustration is directly introduced in
the dynamics. In this case the free energy landscape can be very simple
but only certain transitions between configurations in phase space are
allowed. These models are known under the name of Constrained Kinetic
Models (CKM) \cite{Palmer}, a nice example being the $n$-spin facilitated
Ising model (nSFM) introduced by Fredrickson and Andersen
\cite{FrAn}.

There are few studies (theoretical as well as numerical) of this simple
model but we think it contains some of the fundamental processes
observed in real glasses.
We will study the dynamical properties of one of the simplest models
belonging to the aforementioned second class. In particular we will
concentrate in the 1SFM (to be defined below) at finite dimensions.  We
have observed that there exists a characteristic time $t_c$, independent
of the dimensionality of the system, below which relaxation is
exponential and aging is absent and above which the relaxation becomes
non-exponential and aging appears. This critical time follows an
Arrhenius law with with the temperature and suggests a connection with
some fast processes recently observed by the experimentalists
\cite{Col}.

The paper is organized as follows. In the next section, we define the
1SFM and the main observables we are interested in. Section three
contains some exact relations for one-time staggered quantities at any
dimension. This reveals the existence of two fast processes. Section
four presents numerical simulations in the equilibrium regime and also
in the off-equilibrium regime which evidence the existence of these fast
processes in the two-times correlation and integrated response
function. After the conclusions we present in the Appendix the exact
zero-temperature solution of the 1SFM in one dimension.

\section{The $1SFM$ model: Definition and Observables}

Let us take a set of field variables $\phi(x)$ in a lattice of dimension
$D$ and a Hamiltonian $H\lbrace \phi(x)\rbrace$. Let us consider an
observable $O(t)$ which depends on time $t$ through the configuration of
the system $O(\lbrace\phi_t(x)\rbrace)$. Now we define a discrete time
dynamics for this system. In what follows we will consider a discrete
Monte Carlo (MC) dynamics with random updating. A point of the lattice
is randomly selected and a change of the variable $\phi(x)$ is
proposed. The rate variation of $O$ in $N$ elementary moves (one Monte
Carlo step (MCS)) is given by,

\be
\frac{\partial O(t)}{\partial t}=\overline{P(\phi_t(x,t))
W(\phi_t(x,t)\to\phi'_t(x,t))\Delta O(t)}
\label{eq1}
\ee

where $\Delta O(t)=O(\phi'_t(x))-O(\phi_t(x))$ is the change in an
elementary move of the set of fields $\phi_t(x)$,
$\overline{(\cdot\cdot\cdot)}$ stands for the average over all the
possible transitions and $P(\phi_t(x))$ is the probability of the
configuration $\phi(x)$ at time $t$.

We consider transition probabilities $W$ of the form,

\be
W(\phi_t(x)\to\phi'_t(x))\propto
Min\Bigl (\exp\bigl (-\beta[H(\phi'_t(x))-H(\phi_t(x))]\bigr ),1\Bigr )
\alpha(\phi_t,\phi'_t)
\label{eq2}
\ee

where the term $\alpha(\phi_t,\phi'_t)$ is temperature independent and
cannot, in general, be absorbed in the energy function (note that the
term $\alpha(\phi_t,\phi'_t)$ can be zero for certain transitions while
this can never be the case at finite temperature for the MC
dynamics). In the simplest case where $\alpha=1$ we recover the usual
Metropolis algorithm for the Monte Carlo dynamics. Our purpose is to
study a simple model where frustration only appears in the dynamics via
the term $\alpha(\phi_t,\phi'_t)$ due to the fact that some transitions
between configurations are forbidden.

A simple model of this type is given by the $n$-spin facilitated Ising
model (nSFM) in $D$ dimensions \cite{FrAn}.  To each node of the lattice
we attach a spin variable $\sigma_i$ which can take the values
$0,1$. The energy of the system is defined by the number of spins with
value equal to one (with a minus sign), $E=-\sum_{i}\sigma_i$. By
defining the new set of variables $s=2\sigma-1$ we recognize in the
previous expression the energy of an Ising paramagnet in a magnetic
field $h=\frac{1}{2}$. Fredrickson and Andersen proposed a constrained
dynamics for the nSFM in the following way: a randomly chosen spin in
the lattice is selected and a flip of that spin is proposed
($\sigma_i\to\sigma'_i=1-\sigma_i$). This change is accepted if at least
$n$ of its nearest neighbours are 0 and according to the probability
$Min(e^{-\beta\Delta E},1)$.  The interest of this model is based on the
fact that the non trivial (and glassy) dynamics is all contained in the
term $\alpha(\sigma(t),\sigma'(t))$ (note that the Hamiltonian of this
model has no interaction). Consequently, the dynamics of the nSFM is
highly complex at infinite temperature\footnote{Note that at infinite
temperature the system performs a non trivial random walk in phase
space} while in spin-glasses and other glassy models (where
$\alpha(\phi_t,\phi'_t)$=1) the dynamics in this limit is trivial.  The
nSFM model has been mainly studied in the case $n=1,2$
\cite{FrAn,altres}. In this work we are mainly interested in the case
$n=1$ where some analytic results can be obtained.

There are several physical interpretations of the nSFM
\cite{altres}. The simplest one relates spin variables to the local
compresibility of a fluid region. In this case
regions with very high compressibility can facilitate the mobility of
the neighboring ones while regions of low compressibility lead to a
jamming of the dynamics. 

We define the set of variables $\tau_i=1-\sigma_i$. In terms of this set
the transition probability $W$ eq.(\ref{eq2}) for the 1SFM model
reads \cite{FrAn}.  

\be 
W(\s_x\to\s'_x)\propto \frac{1}{2D}\exp(-\beta\s_x)\sum_{\mu=1}^D
(\tau_{x+e_{\mu}}+\tau_{x-e_{\mu}})
\label{eq3}
\ee

i.e. the transition probability at point $x$ depends linearly on the
magnetisation of the nearest neighbours. The set $\lbrace
e_{\mu};\mu=1,..,D\rbrace$ is a base for the the $D$-dimensional
lattice. This transition probability satisfies detailed balance and
is expected to generate an irreducible Markov process
(in the termodynamic limit) in case $n=1$ for any dimension at 
non-zero temperature (see the reference \cite{altres} for a
discussion on this point).

While the thermodynamics of this model is trivial, its dynamics is much
complicated and only partial results can be obtained in some cases,
especially in one dimension. In this last case, the full dynamics can be
exactly solved at zero temperature. Because we are not aware of this
result in the literature we present it in the Appendix. Unfortunately we
have not been able to close the dynamical equations at finite
temperature.

The general dynamical equation for an observable $O(t)$ in the 1SFM
model is given by 
\be
\frac{\partial O(t)}{\partial t}=\frac{1}{2D}
\Delta O(x) (-e^{-\beta}+(1+e^{-\beta})\tau_x)
(\sum_{\mu=1}^D (\tau_{x+e_{\mu}}+\tau_{x-e_{\mu}}))
\label{eqO}
\ee

\noindent
where $\Delta O(x)=O(\sigma_x=1)-O(\sigma_x=0)$ stands for an elementary
variation of the observable $O$ at time $t$ for the change of $\sigma_x$
from $0$ to $1$. Furthermore, the right-hand side has to be averaged
over all points $x$ of the lattice.

We are interested in the energy-energy correlation function $C_E(t,t')$
$(t'<t)$ defined as,

\be
C_E(t,t')=\frac{\frac{1}{N}\sum_{i=1}^{N}\tau_i(t)\tau_i(t')
-m_{\tau}(t')m_{\tau}(t)}{m_{\tau}(t')(1-m_{\tau}(t'))}
\label{eq4}
\ee

\noindent
where $m_{\tau}(t)=\frac{1}{N}\sum_{i=1}^{N}\tau_i(t)$ is the global
magnetisation associated to the set of variables
$\lbrace\tau_i;i=1,..,N\rbrace$\footnote{The same correlation function
is obtained in terms of the variables
$\lbrace\sigma_i;i=1,..,N\rbrace$}.  
$C_E$ is normalized in such a way that $C_E(t,t)=1$.

We define the staggered one-point and two-point functions,

\bea 
C_0(t)=\frac{1}{N}\sum_{i=1}^{N}\,(-1)^{(\sum_{\mu=1}^{D}i_{\mu})}\,\tau_i\\
C_1(t)=\frac{1}{ND}\sum_{\nu=1}^{D}\sum_{i=1}^{N}\,
(-1)^{(\sum_{\mu=1}^{D}i_{\mu})}\,\tau_i\tau_{i+e_{\nu}}
\label{eq5}
\eea

\noindent
where $(i_{\mu};\mu=1,..,D)$ are the coordinates of the point $i$ in the
$D$-dimensional lattice.  The main interest of defining these quantities
is that they can be exactly closed in any dimension (see next section). 

Another quantity of interest is the energy response function 
$G_E(t,t')$ defined by,

\be
G_E(t',t)=\frac{\delta m(t)}{\delta \beta(t')}~~~~~~~~~~t'<t
\label{resp}
\ee

This function measures the change of energy of the system at time $t$ 
if a small temperature change is done at $t'<t$. Eventually we will be
also interested in the integrated response
function, $IRF(t',t)$ defined as

\be
IRF(t',t)=\int_{t'}^{t}G_E(t'',t) dt''
\label{integrated}
\ee

In order to study the response of the system to a change of temperature
it will be convenient to stay in the linear response regime. In this
regime the change in energy is linear with the
variation of $\beta$. If the largest relaxation time
(associated to the correlation function $C_E(t',t)$) is strongly
dependent with the temperature and if the perturbation of $\beta$ is not
too small then we can expect strong deviations from the
linear response regime. We will return to this point later.


%
%

We should note that staggered functions can be defined only in finite
dimensions and have no meaning in a mean-field version of the model. 
Our main interest is to show that some general results on the 
dynamics of the 1SFM model can be inferred from the staggered one and
two-point functions even though a complete solution of its dynamics at
finite temperature is still lacking.

\section{Fast processes in finite dimensions}

In the finite-dimensional case some general results can be derived for the
staggered magnetisations eq.(\ref{eq5}). In the specific case of the
1SFM the time evolution equations for $C_0(t),C_1(t)$ exactly close
without the need of introducing hierarchies. Using eq.(\ref{eqO}) it is
easy to check that the time evolution equation for $C_0(t)$ reads,

\be
\frac{\partial C_0(t)}{\partial t}=-e^{-\beta}\, C_0(t)
\label{eq13}
\ee

For $C_1(t)$ the following result is obtained, 

\be
\frac{\partial C_1(t)}{\partial t}=-(1+e^{-\beta})\, C_1(t)
\label{eq14}
\ee

This simple result is a consequence of the particular dynamics for the
1SFM as defined in eq.(\ref{eqO}). In case of the 2SFM
\cite{altres} it is not clear if such type of relation exists.
Unfortunately we have not been able to generalize this closure of
equations in case of higher order correlation functions. It is important
to note that staggered correlation functions are zero for homogeneous
conditions (uniform or random for instance). Hence they cannot relax in
time if initially they are at their equilibrium values.  But it is
reasonable to expect the response of the quantities $C_0$ and $C_1$ to a
staggered magnetic field to be also of exponential type.

From previous equations (\ref{eq13}) and (\ref{eq14}) we observe the
existence of at least two characteristic time scales.  One of these time
scales is $t_0=1/(1+e^{-\beta})$, the other one is $t_c=e^{\beta}$. The
time scale $t_0$ does not diverge at zero temperature and is nearly
independent of the temperature ($t_0=\frac{1}{2}$ at infinite
temperature and $t_0=1$ at zero temperature). This time scale
corresponds to the relaxation time for the Ising paramagnet in absence
of interaction. Hence, the time scale $t_0$ corresponds to the
relaxation of a single spin in the presence of a heat-bath and is
independent of the nature of the constrained dynamics.

The time scale $t_c$ is more interesting. It diverges at zero
temperature and has an Arrhenius behavior. Moreover, this time scale
corresponds to the smallest relaxation time of one spin interacting with
its nearest-neighbors and is independent of the dimension of the
lattice. Note that this diverging time scale is a direct consequence of
the constrained dynamics in the system. Then, we conclude that $t_c$ is
the first relevant time scale as a consequence of the cooperative
phenomena which takes place in the 1SFM .  We will see later that this
time scale is associated to a new fast relaxation process.

In addition to these two time scales there is at least another one
$t_{eq}$ associated to the large time decay of the equilibrium two-times
correlation function eq.(\ref{eq4}). This is the largest time scale and
would correspond the $\alpha$-relaxational process observed in glasses
\footnote{Wether in the 1SFM there is also the $\beta$ relaxational
processes as predicted by the Mode Coupling Theory \cite{glasses} is
still unclear} Diagrammatic approximations done by Fredrickson and
Andersen \cite{FrAn} for the two-times correlation function
eq.(\ref{eq4}) show that this time scale coincides with $t_c$. Our
numerical investigation (in agreeement with previous work
\cite{Montero}) suggests that $t_{eq}$ is much larger than $t_c$ and
increases with $\beta$ faster $e^{\beta}$. But more detailed investigations
are necessary in order to a better understanding of this point.

It is natural to think that the exponential relaxation processes
described by equations (\ref{eq13}) and (\ref{eq14}) could be present
also in the relaxation of non-staggered two times quantities like the
energy-energy correlation function eq.(\ref{eq4}) and the integrated
response function eq.(\ref{integrated}). Even though we do not have a
rigorous proof of this assertion we have performed Monte Carlo numerical
simulations for large lattices which clearly show that this is indeed
the case. In what follows, we will focus our research in the physical
consequences of the fast relaxation process described by $t_c$. We will
see the existence of a fast process of exponential character which
manifests in equilibrium and off-equilibrium measurements. Unless
otherwise stated, all results will refer to the 1SFM case.

\section{Numerical results}

\subsection{Numerical algorithm}

We have performed Monte Carlo numerical simulations of the dynamical
equations (\ref{eq3}) in one and two dimensions. Starting from an
initial configuration, the spins in the lattice are randomly selected
and flipped according to the probability eq.(\ref{eq3}). Simulations
were done for relatively large lattice sizes $N=L^D$ with $L=32000$ in
one dimension and $L=200$ in two dimensions with periodic boundary
conditions. In the range of times we are interested in (short time
processes) finite-size corrections are certainly negligible and we have
checked this is really the case.  In order to test our previous exact
results eq.(\ref{eq13}) and eq.(\ref{eq14}), we show in the figure 1 the
exponential decay of $C_0(t)$ at different temperatures as a function of
the rescaled time $t'=te^{-\beta}$ in one dimension starting from the
periodic condition $11101110...$.

\subsection{Equilibrium results}

In order to investigate the equilibrium properties, we should start from
an initial fully thermalized configuration. There are several arguments
which show that a dynamical transition is absent in the 1SFM in finite
dimensions \cite{altres}.  In this case, an equilibrium configuration can
be easily built because the model has no interaction in the energy
function, i.e. the correlation length is zero in equilibrium. We start
from a random initial configuration with energy equal to the equilibrium
energy, the system is let evolve and we compute the correlation function
$C_E(t,t')$. At equilibrium, the $C_E(t,t')$ is time-translational
invariant and depends only on the difference of times
$C_E(t,t')=C_E^{eq}(t-t')$. We have carefully checked this point. We
have repeated this measurements at different temperatures.  We have
observed that correlation functions display two regimes separated by the
critical time $t_c=e^{\beta}$. In the first initial regime (for times
smaller than the critical time $t_c=e^{\beta}$) correlation functions
decay exponentially fast. Empirically we find that the relaxation time
associated to this exponential process is (with very high precision)
$2t_c$,

\be
C_E^{eq}(t)\simeq exp(-t/(2 t_c))=(C_0(t))^{\frac{1}{2}}~~~;~~~t<<t_c
\label{eq15}
\ee

In the second regime, for times larger than $t_c$, the relaxation turns
out to be much slower with stretched exponential behavior at high
temperatures which becomes close to a power law behavior at low
temperatures. We think that no conclusive relaxation behavior can be
guessed numerically in this second regime. A crude estimate of the
relaxation time $t_{eq}$ for the second regime can be obtained
(integrating the equilibrium correlation function respect to the time)
and we obtain values larger than $t_c$. This is an interesting point
which deserves a more detailed investigation and should yield the
relaxation time $t_{eq}$ for the slowest relaxational
process\footnote{Stretched exponential behavior already appears in the
Glauber dynamics of the ferromagnetic one-dimensional Ising model. In
this case it is difficult to guess the correct relaxational behavior
from numerics. See the discussion of Brey and Prados on this
point\cite{BrPr}.}. Figure 2 shows $C_E^{eq}(t)$ in one dimension at
three diferent temperatures $T=0.15,0.2,0.40$. The values of the
critical time associated to these temperatures are $t_c\simeq 780, 150,
12$ respectively. These critical times are indicated with an arrow in
the figure.  The initial exponential decay eq.(\ref{eq15}) is also
depicted in the figure as a continuous line.

Similar results have been obtained in two dimensions where we expect the
this fast process to be independent on the dimension.  This is shown in
figure 3 where we plot the $C_E^{eq}(t)$ for two temperatures
$T=0.15,0.25$ in one and two dimensions and the expected relaxation
eq.(\ref{eq15}). The values of the critical time associated to these
temperatures are $t_c\simeq 780, 55$ respectively. These times are
indicated with an arrow in the figure.  From this figure it can be
clearly appreciated that the regime $t<t_c$ (the region of small times
limited by the arrow) is independent of the dimension of the system as
argued previously. The empirical relaxation eq.(\ref{eq5}) which fits
pretty well the fast decay process is also depicted by a continuous
line.  In the second regime $t>t_c$ relaxation becomes dependent on the
dimension and slower in one dimension than in two dimensions as expected
(dynamical constraints in the 1SFM tend to forbid less paths in phase
space as the dimensionality of the lattice increases).

We have also analyzed the integrated response function at
equilibrium $IRF(t',t)=IRF^{eq}(t-t')$. 
In order to measure it we have prepared the system at equilibrium at a
temperature $\beta$. After some time we change the temperature 
of the system by the quantity $\Delta \beta$ and we let the system evolve 
at the new constant temperature $\beta+\Delta \beta$. If the change of
temperature was applied at zero time then we have,

\be
IRF^{eq}(t)=\frac{m_{\beta+\Delta\beta}(t)-m^{eq}_{\beta}}{\Delta\beta}
\label{eq16}
\ee

\noindent In order to be in the linear response regime it is necessary to
make the change $\Delta\beta$ small enough. But it is important to note
that $\Delta\beta$ cannot be arbitrarily small, otherwise the response
of the system to the temperature change is very small and the
$IRF^{eq}(t)$ measurements become too much noisy. 

In figure 4 we plot the $IRF^{eq}(t)$ normalized to its infinite time
equilibrium value $IRF^{eq}(\infty)$.  This quantity converges to $1$ in
the inifinite time limit. Numerical experiments are shown in one
dimension at two different temperatures $T=0.4$ and $T=0.25$ (data has
been averaged over 20 different runs), for two different signs of the
perturbation $\Delta\beta$. The results of the figure display some
deviations from the linear response regime specially at large times and
low temperatures.  As
commented previously, the equilibrium relaxation time $t_{eq}$ is
strongly dependent on the temperature and increases with $\beta$ faster
than does $t_c$. Then we expect the response of the system to a
perturbation of the temperature to display departures from linearity in
the part of the relaxation process controlled by the time scale
$t_{eq}$. In the part of the relaxation process dominated by $t_c$ this
effects should be smaller. Support in favour of this argument is shown
in the figure 4 where the arrow indicates the value of the critical time
for the two temperatures.  The dependence on the sign of the
perturbation seems to be stronger for the slowest part of the relaxation
(i.e. for times $t>t_c$) than in the fast part of the relaxation
(i.e. for times $t<t_c$). In order to get more clear cut results it
would be necessary to go to lower temperatures. Unfortunately the
$IRF^{eq}(t)$ becomes too much noisy even though we have been able to
confirm this trend.

\subsection{Off-equilibrium results}

In this section we want to investigate the role of the critical time
$t_c$ for non equilibrium relaxation processes. We are going to show
that $t_c$ sets a minimum time scale above which non-equilibrium
relaxations display aging effects. In the following we will use the
notation in which $t'\to t_{w}, t\to t_w+t$ ($t_w$ stands for waiting
time).  We have done numerical simulations of the 1SFM in one and two
dimensions measuring $C_E(t_w,t_w+t)$.

Results for the $C_E(t_w,t_w+t)$ are shown in figure 5 in the one
dimensional case. We start from a random initial condition\footnote{The
same results were obtained starting from a periodic initial
configuration like for instance $11101110...$.  where $C_0(t=0)$ and
$C_1(t=0)$ are different from zero.} with initial energy $E(t=0)=-0.5$
quite far from its equilibrium value
($E^{eq}(\beta)=-1/(1+e^{-\beta})$). We computed the $C_E(t_w,t_w+t)$
for different values of the waiting time $t_w=10,10^2,10^3,10^4,10^5$ at
$T=0.1$ (at this temperature $t_c\simeq 22000$) in one and two
dimensions (we show only the results in one dimension, in two dimensions
they are qualitatively similar). In order to clearly appreciate the
qualitative trend of the data we only show the values of $t$ within the
range $10^3-10^5$ MCS.  Figure 5 shows two regimes. In the first regime
($t_w=10,10^2,10^3<t_c$) aging effects are absent, i.e. the relaxation
curve $C_E(t_w,t_w+t)$ only depends on $t$. In the regime $t_w=10^5 >
t_c$ aging effects appear (i.e. the full relaxation curve
$C_E(t_w,t_w+t)$ also depends on $t_w$). The continuous line is the
equilibrium exponential behavior eq.(\ref{eq15}).


As emerges from figure 5 the aging effects in the correlation funtion
are very small even for $t_w>t_c$ at least in the region of times where
the value of the correlation function is not too small. In fact, a
scaling of the type $C_E(t_w,t_w+t)=f(t/t_w)$ will not wotk at all. This
is surprising since one would expect in the off-equilibrium slow regime
a dynamical behavior plagued of strong aging effects in the correlation
function. Possibly, this is a consequence of the particular correlation
function used. In what follows we will see that the response function
displays clear aging effects.

We have measured the off-equilibrium integrated response function
starting from a random initial configuration. The system is let evolve
for a time $t_w$ at constant (inverse temperature) $\beta$.  At time
$t_w$ we make a copy of the system and we make it evolve at the constant
new temperature $\beta+\Delta\beta$. After $t_w$ we measure the
difference between the magnetisations of the two copies evolving with
the same thermal noise but different temperatures.  In this way we
compute

\be
IRF(t_w,t_w+t)=\frac{m_{\beta+\Delta\beta}(t+t_w)-
m_{\beta}(t+t_w)}{\Delta \beta}
\label{eq17}
\ee

In figure 6 we show the normalized integrated response function
$IRF(t_w,t_w+t)/IRF^{eq}(\infty)$ at $T=0.1$ ($t_c\simeq 22000$) for
different values of $t_w=10,10^2,10^3,10^4,10^5$ in the one dimensional
case.  We can clearly appreciate the existence of the critical time
$t_c$ which separates a regime where there is no $t_w$ dependence from a
regime where the system {\em ages}.  As $t_w$ increases the response of
the system becomes smaller which is a typical feature of aging
\cite{Bou}. Note that the equilibrium value of $IRF^{eq}(t)$ in the
window of time shown in the figure 6 is practically zero (and converges
to 1 for large times) while in the
off-equilibrium regime the response reaches a (negative!) aproximate
value of $-10^3$. Hence the response of the system is three orders of
magnitude larger and negative in the off-equilibrium regime than in the
equilibrium case.

 

\section{Conclusions}

This work has been devoted to the study of one of the simplest
constrained kinetic Ising models, i.e. the so called nSFM introduced by
Friedrickson and Andersen in the special case of $n=1$. This is an
interesting model with an extremely simple energy
function where the frustration is contained in the dynamical rules which
forbid certain transitions between configurations in phase space.

Our main interest has been the research of fast processes in the
1SFM. In this model the rate of variation of the energy in one-point of
the lattice is linearly coupled to the energy of the nearest
neighbors. This is the simplest case one can consider, while the 2SFM
corresponds to a quadratic coupling between the nearest neighbors sites
in the lattice\cite{FrAn}. In case of the 1SFM some exact closed
dynamical equations can be obtained for the one and two-point staggered
functions $C_0$ and $C_1$ in any dimensions. This reveals that both
staggered functions $C_0$ and $C_1$ decay exponentially fast (with
characteristic times $t_c$ and $t_0$) independently of the
dimensionality of the lattice. $t_0$ is the relaxation of a single spin
uncoupled to its nearest neighbors embedded in a thermal bath (hence,
independent of the dimension) and $t_c$ is an Arrhenius temperature
dependent relaxation time resulting from the dynamical constraints. Then
we expect the existence of these fast processes in other non-staggered
quantities like the energy-energy correlation function and also the
integrated response function.  While we have not found a precise
demonstration of this result we have given strong numerical support to
this hypothesis by measuring the equilibrium and non-equilibrium
behavior of the correlation function eq.(\ref{eq4}) and integrated
response function eq.(\ref{integrated}) in one and two dimensional
lattices.  In the equilibrium case the correlation function
eq.(\ref{eq4}) decays exponentially fast with an (empirically found)
relaxation time $2t_c$ in the regime of times $t<t_c$ while relaxation
becomes slower in the regime $t>t_c$. Furthermore, the integrated
response function displays strong non-linear effects in the regime
$t>t_c$ specially at low temperatures. Concerning the off-equilibrium
behavior we find that aging effects are absent for values of the waiting
time less than the critical time. This is nicely observed in the
behavior of the integrated response function (figure 6).

Now we should discuss to what extent our results are general and not
confined to the particular 1SFM case. The equations (\ref{eq13}) and
(\ref{eq14}) are exact results in the 1SFM case and we expect they are
not more valid in the nSFM with $n$ larger than $1$. We have performed
some numerical simulations in case of the 2SFM but we have not found
evidence on the existence of this critical time. This could explain why
some previous numerical works on the 2SFM \cite{altres} were able to fit
the equilibrium relaxation functions to a strecthed exponential behavior
while in the 1SFM this is not possible due to the existence of two
different regimes. In numerical simulations one is able to explore only
relatively small scales of time. It is clear that an initial
exponential process would make numerical fits of the slow regime
difficult, specially in the low temperature phase where the critical time
starts to become large.

Also in the realm of disordered systems (for instance in spin glasses
\cite{Books}) we are not aware of the existence of this critical time,
possibly because in that case frustration is directly introduced in the
energy function and not in the dynamics. We are tempted to conclude that
(1) the existence of a temperature activated critical time and (2) the
presence of exponential decay processes in the relaxation of some
correlation functions in the short-time regime, are both strictly
related to the nature of this type of constrained short-range dynamics.

Summarizing, we have found evidence about the existence of two
fast processes in the 1SFM model. The first process with
characteristic time $t_0$ corresponds to the relaxation of single spin
in the lattice. This is a trivial process not related to any cooperative
effect in the lattice. The second process is a fast exponential process
consequence of the dynamical constraints. Physically it would correspond to
the relaxation of some coupled degrees of freedom of the system.  The
results which emerge from the study of the 1SFM are in agreement with
some recent experimental findings by Colmenero et al. on fragile polymer
glasses.  By doing neutron scattering measurements in glasses they claim
on the existence of a temperature activated critical time (with small
energy barrier) which separates two well defined time regions. In the
first region the relaxation is exponential while in the second regime
relaxation is much slower with higher relaxation time. This feature
seems to be captured by the present model.

It would be very interesting to analitically solve the dynamics at
finite temperature of this model (at least in one dimension) in order to
confirm the numerical findings of this work. This would also shed light
on the existence of other relaxational processes (like the $\alpha$ and
$\beta$ relaxation) as predicted by the Mode Coupling Theory.

\begin{center}
{\bf Acknowledgements}
\end{center}
F.R acknowledges Ministerio de Educacion y Ciencia of Spain for
financial support. E. F. acknowledges Gobierno de Navarra for financial
support through a predoctoral grant. We are grateful to Vicente Azcoiti
and Paco Padilla for a careful reading of the manuscript.

\section{Appendix}

In this appendix we present the exact solution of the 1SFM in one
dimension at zero temperature. Lets take a chain of $N$ spins $\s_i$ and
we define the new set of variables $\tau_i=1-\s_i$. In terms of this set
of variables we define the following set of correlation functions,

\be
D_{k}(t)=\frac{1}{N}\sum_{r=1}^N\tau_r(t)\tau_{r+1}(t)\,...\,\tau_{r+k}(t)
\label{ap1}
\ee

The magnetisation $m_{\tau}=\frac{1}{N}\sum_{r=1}^N\tau_r=D_0(t)$ is the
first term of this hierarchy. Using eq.(\ref{eqO}) we can derive the
time evolution of the $D_k(t)$ at zero temperature. We get,

\be
\frac{\partial D_{k}}{\partial t}=-\,D_{k+1}-\,k\,D_k;~~~~~k\ge 0
\label{ap2}
\ee

We introduce the generating function 

\be
G(x,t)=\sum_{k=1}^{\infty}\frac{x^k}{k!}D_k(t)
\label{ap3}
\ee

In terms of this generating function we have the following partial
differential equation,

\be
\frac{\partial G(x,t)}{\partial t}=-(1+x)\frac{\partial G(x,t)}{\partial x}
\label{ap4}
\ee

This linear partial differential equations is readily solved yielding,

\be
G(x,t)=G_0((1+x)e^{-t}-1)
\label{ap5}
\ee

where $G_0(x)=G(x,0)$ is the initial condition. We can obtain the
different set of moments, 

\be
D_k=\Bigl (\frac{\partial^k G}{\partial x^k}\Bigr )_{x=0}
\label{ap6}
\ee

In particular we get for the magnetisation
$m_{\tau}=G_0(\exp(-t)-1)$. In the large time limit it converges to
$G_0(-1)$ depending on the initial condition. For the particular initial
condition $\s_i=0$, $G_0(x)=e^x$ and the magnetisation $m_{\tau}$ does
not converge to its equilibrium value ($m_{\tau}^{eq}=1$) but to $1/e$.

We can also compute the two-times correlation function
eq.(\ref{eq4}). At zero temperature we find, 

\be
C_E(t',t)=\frac{1-m(t)}{1-m(t')}=\frac{m_{\tau}(t)}{m_{\tau}(t')}=
\frac{G_0(e^{-t}-1)}{G_0(e^{-t'}-1)}
\label{ap7}
\ee

In figure 7 we compare previous equation (\ref{ap7}) with the numerical
results for different values of $t'$.

\vfill
\newpage
{\bf Figure Captions}
\begin{itemize}

\item[Fig.~1] Exponential relaxation of $C_0(t)$ in one dimension 
for several temperatures $T=0.1,0.2,0.3,0.4,0.5$.

\item[Fig.~2] $C_E^{eq}(t)$ in one dimension at temperatures
$T=0.15,0.25,0.4$. The continuous lines are the exponential relaxations
eq.(\ref{eq15}). 

\item[Fig.~3] $C_E^{eq}(t)$ at temperatures $T=0.15$ ($D=2 squares, D=1
(crosses)$) $T=0.25$ ($D=2 (times), D=1 (circles)$) in one and two
dimensions. The continuous lines are the exponential relaxations
eq.(\ref{eq15}).

\item[Fig.~4] Normalized $IRF^{eq}(t)$ in one dimension at temperatures
$T=0.4$ with $\Delta\beta=0.2 (squares), -0.2 (crosses)$ and
$T=0.25$ with $\Delta\beta=0.4 (rhombs), -0.4 (dots)$

\item[Fig.~5] $C_E(t_w,t_w+t)$ for different waiting times
$t_w=10,10^2,10^3,10^4,10^5$ in one dimension at $T=0.1$. For
$t_w=10,10^2,10^3$ the relaxation curves superimpose and aging is
absent.

\item[Fig.~6] Normalized $IRF(t_w,t_w+t)$ at $T=0.1$ for several values
of $t_w$ and $\Delta\beta=0.5$. The symbols are guide to the eyes. Note
that $IRF^{eq}(t)$ is nearly zero in this range of times.

\item[Fig.~7] $C_E(t_w,t_w+t)$ for different
waiting times $t_w=0.0125,0.125,1.25$ at zero temperature in one dimension. 
The lines are the exact solution (\ref{ap7}).

\end{itemize}


\begin{thebibliography}{99}

\bibitem{glasses} W. Gotze, {\em Liquid, freezing and the Glass transition},
Les Houches (1989), J. P. Hansen, D. Levesque, J. Zinn-Justin editors,
North Holland; C. A. Angell, Science, {\bf 267}, 1924 (1995)

\bibitem{MCT} E. Lentheusser, Phys. Rev.{\bf A29}, 2765 (1984);
T. R. Kirkpatrick, Phys. Rev {\bf A31}, 939 (1985);
W. Gotze and L. Sjogren, Rep. Prog. Phys. {\bf 55}, 241 (1992)


\bibitem{neutron} F. Fujara and W. Petry, Europhys. Lett. {\bf 4}, 571
(1987); B. Frick and D. Richter, Phys. Rev. {\bf B47}, 14795 (1993)

\bibitem{Col} J. Colmenero, A. Arbe and A. Alegria,
Phys. Rev. Lett. {\bf 71}, 2603 (1993)

\bibitem{FrPa} S. Franz and G. Parisi, {\em Recipes for metastable
states} Preprint cond-mat/{\bf 9503167} and references therein

\bibitem{BG} F. Ritort, Phys. Rev. Lett. {\bf 75}, 1190 (1995);
A. Barrat and M. Mezard, J. Physique I (Paris) {\bf 2}, 705 (1992)

\bibitem{Books}  M.~M\'ezard, G.~Parisi and  M.~A.~Virasoro,
{\em Spin Glass Theory  and Beyond}
(World Scientific, Singapore 1987);
K. H. Fischer and J. A. Hertz, {\em Spin Glasses} (Cambridge University
Press 1991);

\bibitem{MaPaRi} T. R. Kirkpatrick and D. Thirumalai, J. Phys. A
(Math. Gen.) {\bf 22}, L149 (1989); J. P. Bouchaud and M. Mezard,
J. Physique I (Paris) {\bf 4}, 1109 (1994); E. Marinari, G. Parisi and
F. Ritort, J. Phys. A (Math. Gen.) {\bf 27}, 7615 (1994); J. Phys. A
(Math. Gen.) {\bf 27}, 7647 (1994);

\bibitem{Palmer} R. G. Palmer in  {\em Lectures in the Sience of
Complexity } Ed. D. L. Stein, Sta. Fe Institute (Addison Wesley 1989)
and references therein; G. H. Fredrickson, Ann. Rev. Phys. Chem. {\bf
39}, 149 (1988)

\bibitem{FrAn} G. H. Fredrickson and H. C. Andersen, Phys. Rev. Lett
{\bf 53}, 1244 (1984)

\bibitem {MFSG} S. Franz and J. Hertz, Phys. Rev. Lett. {\bf 74}, 2114
(1995); J. P. Bouchaud, L. F. Cugliandolo, J. Kurchan and M. Mezard,
{\em Mode-coupling approximations, glass theory and disordered systems}
Preprint cond-mat/{\bf 9411042}; J. P. Bouchaud, A. Comtet and
C. Monthus, {\em On a dynamical model of glasses} Preprint cond-mat/{\bf
9506027} to appear in J. Physique I.

\bibitem{altres}  
G. H. Fredrickson and H. C. Andersen, J. Chem. Phys. {\bf 83}, 5822
(1988); G. H. Fredrickson and S. A. Brawer, J. Chem. Phys. {\bf 84}, 3351 

\bibitem{Montero} M. J. Ruiz-Montero, {\em Modelos sencillos de relajacion
estructural en vidrios} PhD Thesis, Sevilla (November 1992); J. J. Brey and
M. J. Ruiz-Montero, Phys. Rev. {\bf B43} (1991) 585
 
\bibitem{BrPr} J. J. Brey and A. Prados, {\em  Low temperature
relaxation in the one-dimensional Ising model} to appear in
Phys. Rev. E. 

\bibitem{Bou} J. P. Bouchaud, J. Physique I (Paris) {\bf 2}, 1705
(1992); L. F. Cugliandolo and J. Kurchan, Phil. Magaz. {\bf B71}, 50
(1995);
 

\end{thebibliography}
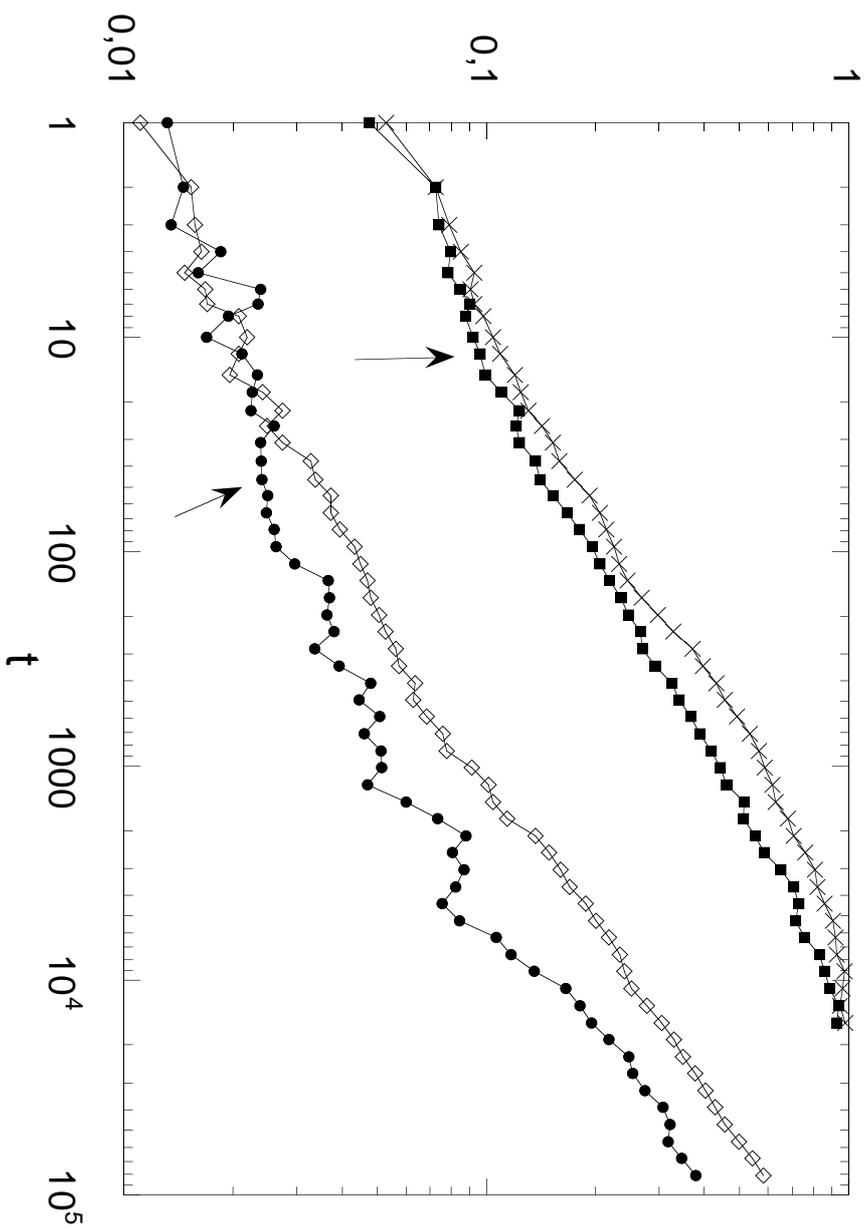
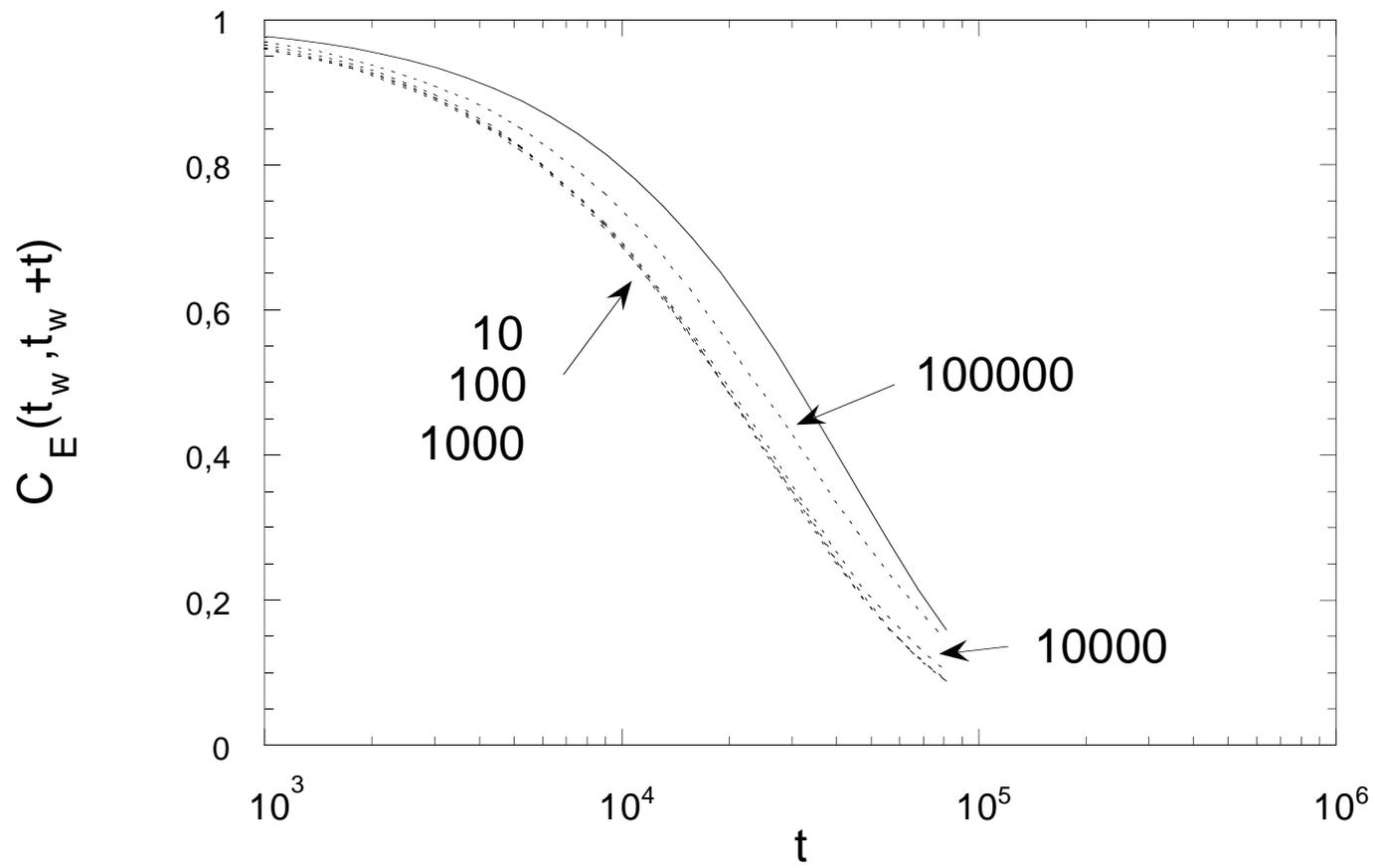
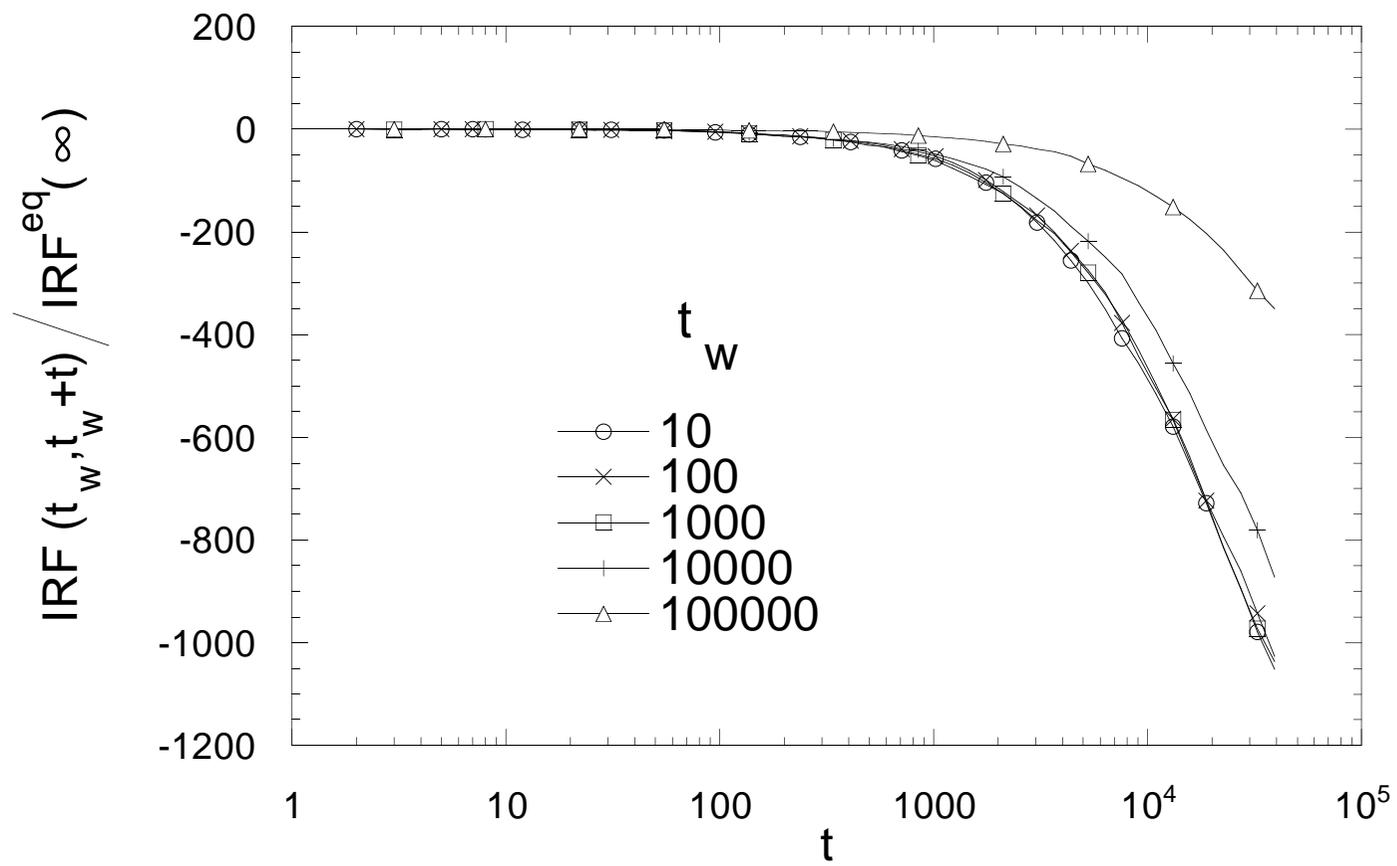
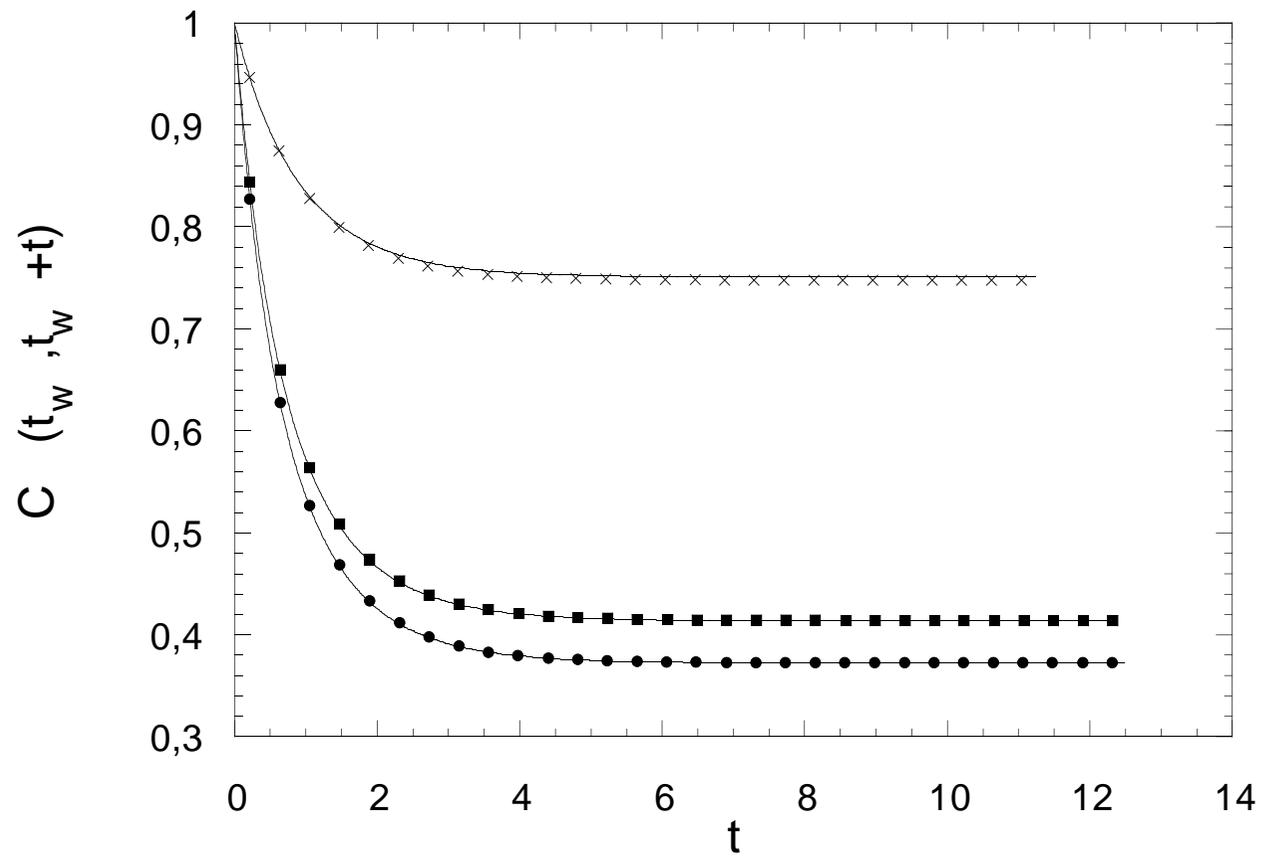
\end{document}